\documentclass[aps,prb,twocolumn,groupedaddress,showpacs]{revtex4-1}
\usepackage[centertags]{amsmath}
\usepackage{graphicx,amssymb}

\begin{document}
\title{Mott correlated states in the underdoped two-dimensional Hubbard model:
 variational Monte Carlo versus a dynamical cluster approximation.}

\author{Luca F. Tocchio}
\affiliation{Institut f\"ur Theoretische Physik, Goethe-Universit\"at Frankfurt, Max-von-Laue-Stra{\ss}e 1, 60438 Frankfurt am Main}
\author{Hunpyo Lee}
\affiliation{Institut f\"ur Theoretische Physik, Goethe-Universit\"at Frankfurt, Max-von-Laue-Stra{\ss}e 1, 60438 Frankfurt am Main}
\author{Harald O. Jeschke}
\affiliation{Institut f\"ur Theoretische Physik, Goethe-Universit\"at Frankfurt, Max-von-Laue-Stra{\ss}e 1, 60438 Frankfurt am Main}
\author{Roser Valent\'\i}
\affiliation{Institut f\"ur Theoretische Physik, Goethe-Universit\"at Frankfurt, Max-von-Laue-Stra{\ss}e 1, 60438 Frankfurt am Main}
\author{Claudius Gros}
\affiliation{Institut f\"ur Theoretische Physik, Goethe-Universit\"at Frankfurt, Max-von-Laue-Stra{\ss}e 1, 60438 Frankfurt am Main}
\date{\today}

\begin{abstract}
We investigate the properties of the frustrated underdoped Hubbard
model on the square lattice using two complementary approaches, the
dynamical cluster extension of dynamical mean field theory, and
variational Monte Carlo simulations of Gutzwiller-Jastrow
wavefunctions with backflow corrections. We compare
and discuss data for the energy and the double occupancies, 
as obtained from both approaches. At small
dopings, we observe a rapid crossover from a weakly correlated metal
at low interaction strength $U$ to a non-Fermi liquid correlated state
with strong local spin correlations.
Furthermore, we investigate the stability of the correlated state against phase
separation. We observe phase separation only for large values of $U$
or very large frustration. No phase separation is present for the
parameter range relevant for the cuprates.
\end{abstract}

\maketitle

\section{Introduction}\label{sec:intro}
The Hubbard model on the square lattice is a minimal model 
for describing electronic correlation. This model
has played a central role for the study of 
high-$T_C$ superconductivity, since it is believed to 
capture the essential physics of the copper-oxygen planes 
in cuprate materials. Furthermore, the properties of the Hubbard 
model in the underdoped regime, {\it i.e.} in the proximity of the Mott 
insulating state at half-filling, is a stimulating research 
area due to the challenge in describing the physics of
the correlated pseudogap state and its
 non-Fermi liquid behavior. 
Significant achievements in this field have been 
made, for instance, by the cluster extensions of dynamical mean-field 
theory~\cite{capone,Sakai2009,civelli,jarrell,gull,sordi,liebsch}, 
by means of a phenomenological theory~\cite{rice} and of the Gutzwiller 
approximation~\cite{markiewicz}, as well 
as on the basis of increasingly accurate variational wave 
functions~\cite{randeria,tocchio_FS}.

Recently, it has been proposed that the pseudogap and the superconducting phases 
present at finite doping~\cite{werner,gull2,sordi2}  and for onsite
$U$ values corresponding to a Mott insulating state at half filling, can be continuously connected 
to a pseudogap and a superconducting phase at half-filling for $U$
values lower  than the critical $U_c$ corresponding to  the 
Mott-Hubbard metal-insulator transition (MIT). This proposal contrasts with
Anderson's concept of superconductivity as a state emerging 
out of a Resonant Valence Bond (RVB) state~\cite{anderson}, a 
prototypical Mott-insulating state at half filling in the absence of 
magnetic order.

Another interesting feature of a correlated electron state in
the underdoped regime is the tendency of the system to phase 
separate into an undoped state with strong antiferromagnetic 
correlations and a hole-doped region. Indeed, if the Hubbard model 
would be unstable to phase separation, its validity as a model  
to properly describe the development of superconductivity 
could be questioned. Phase separation occurs when the stability 
condition $\partial^2 E(n)/\partial n^2 >0$ is violated, 
{\it i.e.} when the ground-state energy $E(n)$, as a function of 
electronic density $n$, is not any more convex.
As introduced by Emery \emph{et al.} \cite{emery}, 
phase separation can be studied by looking at the 
hole energy $E_{\delta}(\delta)$, defined as, 
\begin{equation}\label{Emery}
E_{\delta}(\delta)=\frac{E(\delta)-E(0)}{\delta}\, 
\end{equation}
where $\delta=1-n$ is the hole density. If the hole energy has a
minimum at a critical doping $\delta_c$, the system is unstable to
phase separation for $\delta < \delta_c$.  A difficulty when using the
Emery criterion (\ref{Emery}) is its strong dependence on the accuracy
in the estimate of the ground state energy, as discussed in
Ref.~\onlinecite{becca}.  In particular, less accurate ground state
energies tend to overestimate the critical $\delta_c$ below which the
system exhibits phase separation.  This is a consequence of the fact
that good estimates for the energy of low-doping states are hard to
obtain because of the strong local correlations. Different approaches
have led in the past to contradictory results for the Hubbard model in the parameter range 
relevant for hole-doped cuprates, ranging from absence of phase separation
 to a range of estimates for phase separation up to 10\%
doping~\cite{Maier,aichhorn,chang,chen,trivedi,chen2,watanabe}. For the $t-J$
model~\cite{valenti,lugas,troyer,hu}, {\it i.e.} for the large-$U$
limit of the Hubbard model, phase separation is present at all dopings
for large values of $J$, but the possible occurrence of phase
separation close to half filling for small values of $J$ has been a
long-standing debate~\cite{lugas,troyer,hu}.

In this work, we make use of variational Monte Carlo (VMC)
simulations, including backflow correlations~\cite{backflow}, 
and the dynamical cluster approximation (DCA)~\cite{Hettler1998,Maier2005} 
to address the properties of the
 Hubbard model on the square lattice with nearest and next-nearest
neighbor hoppings in the 
underdoped regime both at zero and finite temperature. 
As already pointed out in Ref.~\onlinecite{tocchio_FS}, 
we distinguish a weakly correlated metal at small $U$ 
({\it i.e.} continuously connected to a half-filled metallic regime) 
and a strongly correlated state 
at intermediate to large $U$ ({\it i.e.} when the half-filled 
case is insulating). We find significant differences between the
 two regimes  in the static and in the dynamical spin correlations 
as well as in the low-energy part of the self-energy. 
 The above observations  obtained from two
complementary approaches, VMC and DCA,  and the violation of
the Luttinger sum rule for the strongly correlated state
reported in a previous study~\cite{tocchio_FS}, 
hint to a non-Fermi liquid nature of the strongly correlated
state. 
The two states are 
separated by a crossover line at the critical
 interaction $U_c$, that may evolve 
into a first-order transition~\cite{sordi} when doping is vanishing. 

 Moreover, we investigate the occurrence of phase separation in the
underdoped Hubbard model. We find no tendency to phase separation when
the Coulomb repulsion $U$ is slightly above the critical $U_c$,
regardless of the ratio $t'/t$ between next-nearest and nearest
neighbor hoppings, $t'$ and $t$ respectively. Our results therefore 
indicate that for intermediate values of the Coulomb repulsion
($U/t\sim 6-8$), the Hubbard model is stable against phase separation,
irrespective of the value of $t'/t$. The actual value of $t'/t$ is
sensitive to the details of the high-$T_c$ cuprate compound
investigated and can be calculated, for example, in density functional
theory, leading to $t'/t\simeq[-0.1,-0.4]$.  A larger degree of
frustration has been shown to correlate, in general, with higher
critical temperatures~\cite{andersen}.

If we increase the electron repulsion $U$ at constant $t'/t$, the
system is, in contrast, found to be unstable towards phase
separation. However, since increasing the variational accuracy tends to
suppress phase separation, we cannot exclude that an even more
accurate approach could eliminate or significantly reduce the tendency
to phase separate also for larger values of the Coulomb repulsion.

The paper is organized as follows: in Section~\ref{sec:method}, we
introduce the Hamiltonian and describe the VMC and the DCA approaches;
in Section~\ref{sec:energy}, we compare ground-state energy and the
density of double occupancies within the two methods, in
Section~\ref{sec:NFL}, we discuss the non-Fermi liquid properties of
the underdoped region at $U>U_c$, in contrast with the weakly
correlated metal at $U<U_c$. In Section~\ref{sec:phase_sep} the
occurrence of phase separation is discussed and finally we present our
conclusions in Section~\ref{sec:concl}.

\begin{figure}
\includegraphics[width=1.0\columnwidth]{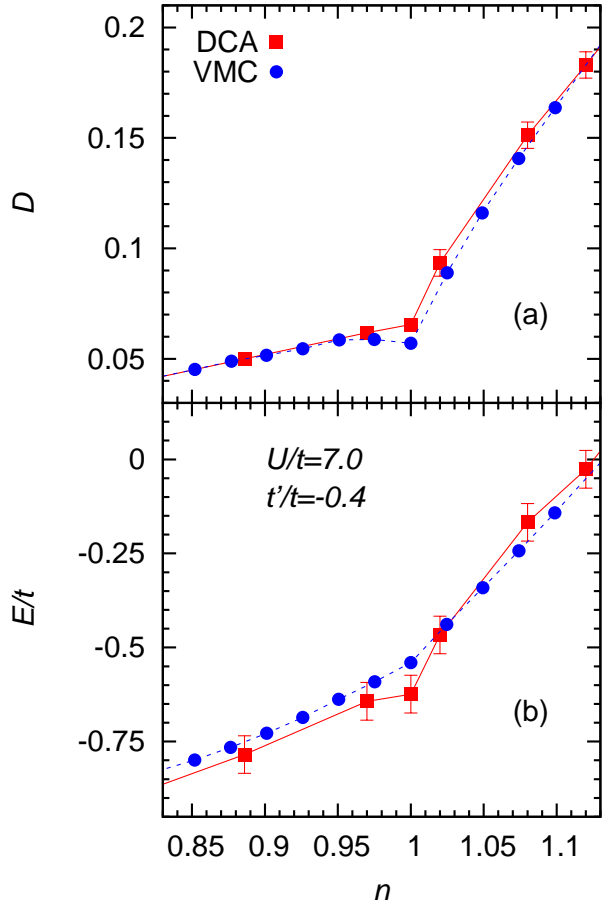}
\caption{\label{fig:energy_com}
(Color online) (a) Double occupancy $D$ and 
(b) energy as a function of electronic density $n$ 
for $U/t=7.0$ and $t'=-0.4t$. Results are obtained using a VMC 
approach on an $L=162$ lattice size (blue circles) and within 
DCA by means of a $2 \times 2$ plaquette in $k$-space 
for a temperature $T/t=0.05$ (red squares). Errors in VMC are smaller than the symbol size. 
Lines are just guides to the eye.}
\end{figure}

\section{Model and methods}~\label{sec:method}

We consider the frustrated Hubbard model with extended 
hopping on a two-dimensional square lattice, 
\begin{equation}\label{eq:hamiltonian}
{\cal H} = -t \hspace{-0.5ex} \sum_{\langle ij\rangle\sigma} 
     \hspace{-1ex} c^{\dagger}_{i\sigma} c^{\phantom{\dagger}}_{j\sigma}
     -t^\prime \hspace{-0.5ex} \sum_{\langle\langle ij\rangle\rangle\sigma} 
     \hspace{-1ex} c^{\dagger}_{i\sigma} c^{\phantom{\dagger}}_{j\sigma}  
     +\textrm{H.c.} 
     +U \sum_{i} n_{i\uparrow} n_{i\downarrow},
\end{equation}
where $c^{\dagger}_{i\sigma}$ ($c^{\phantom{\dagger}}_{i\sigma}$) denotes the 
electron creation (annihilation) operator of one electron on site $i$ with spin 
$\sigma=\uparrow,\downarrow$, $\langle ij\rangle$ and 
$\langle\langle ij\rangle\rangle$ indicate nearest and next-nearest neighbor 
sites respectively; 
$n_{i\sigma}=c^{\dagger}_{i\sigma}c^{\phantom{\dagger}}_{i\sigma}$ is the 
electron density; $t$ and $t^\prime$ are the nearest and next-nearest neighbor
hopping amplitudes, and $U$ is the on-site Coulomb repulsion.

\subsection{Variational Monte Carlo}

Variational Monte Carlo (VMC) simulations allow to perform non-perturbative 
calculations at zero temperature for one- and two-dimensional 
correlated and frustrated systems. VMC simulations are based on
numerically sampling expectation values over a variational 
estimate of the ground-state wave function. Here we will 
use a powerful variational {\it ansatz} 
for frustrated electron systems which has been strictly tested 
by comparing extensively to analytical and numerical exact 
limiting cases. Our variational {\it ansatz} has been tested in
particular against Bethe {\it ansatz} predictions for the
Luttinger liquid exponents in one dimension \cite{capello},
with respect to Lanczos results, in terms of wave function 
overlap, for a two-dimensional 18 site cluster \cite{lanczos},
and with respect to the results of density matrix renormalization 
group studies, in terms of the ground-state energy, for the 1D 
$J_1-J_2$ model \cite{backflow}.

\begin{figure}
\includegraphics[width=1.0\columnwidth]{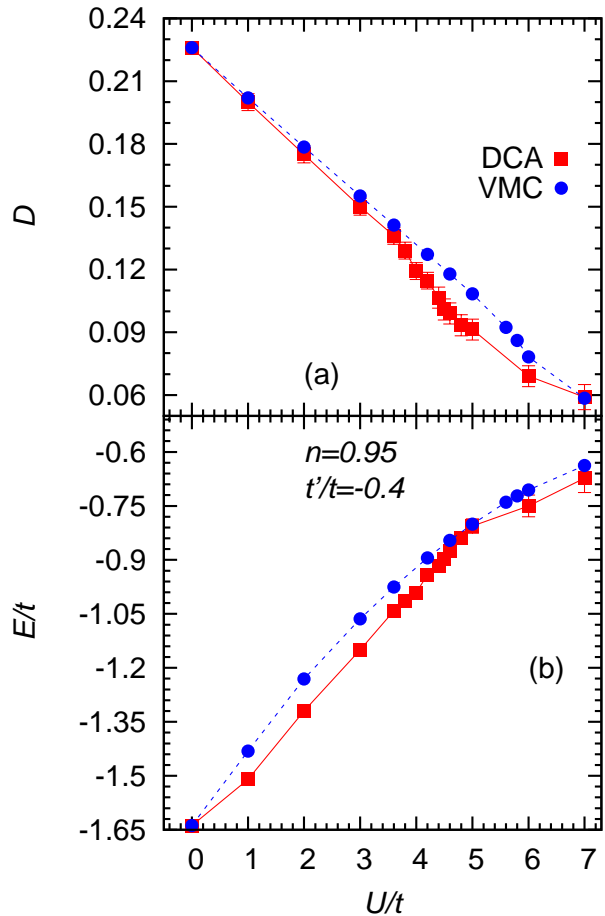}
\caption{\label{fig:Energy} (Color online) (a) Double occupancy $D$
  and (b) energy as a function of $U/t$ for electron density $n=0.95$
  and $t'=-0.4t$. Data are obtained by a VMC approach on an $L=162$
  lattice size (blue circles) and within DCA by means of a $2 \times
  2$ plaquette in $k$-space for a temperature $T/t=0.05$ (red
  squares). Errors in VMC are smaller than the symbol size.  Lines are
  just guides to the eye.}
\end{figure}

The variational {\it ansatz} consists of three components.
In a first step,  we construct uncorrelated non-magnetic wave 
functions given by the ground state $|\rm{BCS}\rangle$ 
of a superconducting Bardeen-Cooper-Schrieffer (BCS) 
Hamiltonian~\cite{grosbcs,zhang,grosAnnals}: 
\begin{equation}\label{eq:meanfield}
{\cal H}_{\rm{BCS}} = \sum_{k\sigma} \xi_k 
c^{\dagger}_{k\sigma} c^{\phantom{\dagger}}_{k\sigma}
+ \sum_{k} \Delta_k 
c^{\dagger}_{k\uparrow} c^{\dagger}_{-k\downarrow} + \rm{H.c.},
\end{equation}
where both the free-band dispersion $\xi_k$ and the pairing amplitudes 
$\Delta_k$ are variational functions. We use the parametrization
\begin{eqnarray}
\xi_k&=& -2\tilde{t}(\cos k_x +\cos k_y)
-4\tilde{t}^\prime\cos k_x \cos k_y -\mu \\
\Delta_k&=& \,2\Delta_{\textrm{BCS}}(\cos k_x -\cos k_y), 
\label{eq:effective} 
\end{eqnarray}
where the effective hopping amplitude $\tilde{t}^\prime$, the
effective chemical potential $\mu$, and the local pairing field
$\Delta_{\textrm{BCS}}$ are variational parameters to be
optimized in order to minimize the variational energy. 
The parameter $\tilde{t}$ is kept fixed to set the energy
scale. We point out that large electronic correlations, as in proximity 
of a Mott insulator, lead to a strong renormalization of $\tilde{t}^\prime$ 
with respect to the bare Hamiltonian value $t'$, see Ref.~\onlinecite{tocchio_FS}.

The correlated state $|\Psi_{\textrm{BCS}}\rangle$, without backflow
terms, is then given by $|\Psi_{\textrm{BCS}}\rangle = {\cal J}
|\textrm{BCS}\rangle$, where ${\cal J}=\exp(-1/2 \sum_{ij} v_{ij} n_i
n_j)$ is a density-density Jastrow factor (including the on-site
Gutzwiller term $v_{ii}$), with the $v_{ij}$'s being optimized for
every independent distance $|i-j|$. Notably, within this kind of wave
function, it is possible to obtain a pure ({\it i.e.}, non-magnetic)
Mott insulator for a sufficiently singular Jastrow factor $v_q \sim
1/q^2$ (where $v_q$ is the Fourier transform of $v_{ij}$), while a
superconducting (metallic) state is found whenever $v_{q}\sim 1/q$ and
$\Delta_{\textrm{BCS}}>0$ ($\Delta_{\textrm{BCS}}=0$)~\cite{capello}.

A size-consistent and efficient way to further improve the correlated
state $|\Psi_{\rm{BCS}}\rangle$ for large on-site interactions is
based on backflow correlations. In this approach, each orbital that
defines the unprojected state $|\textrm{BCS}\rangle$ is taken to
depend upon the many-body configuration, in order to incorporate
virtual hopping processes~\cite{backflow}, in particular the recombination 
of neighboring charge fluctuations is favored. This is a substantial improvement 
with respect to Jastrow factors, where electron-electron correlation 
is included only via a multiplicative term. All results presented here
are obtained by fully incorporating the backflow corrections and
optimizing individually~\cite{stoc_ref} every variational parameter in
$\xi_k$ and $\Delta_k$, in the Jastrow factor ${\cal J}$, as well as
for the backflow corrections. Calculations are performed on
$45^{\circ}$ tilted clusters with $L=162$ lattice sites and periodic
boundary conditions.

\subsection{Dynamical Cluster Approximation}

\begin{figure}
\includegraphics[width=0.8\columnwidth]{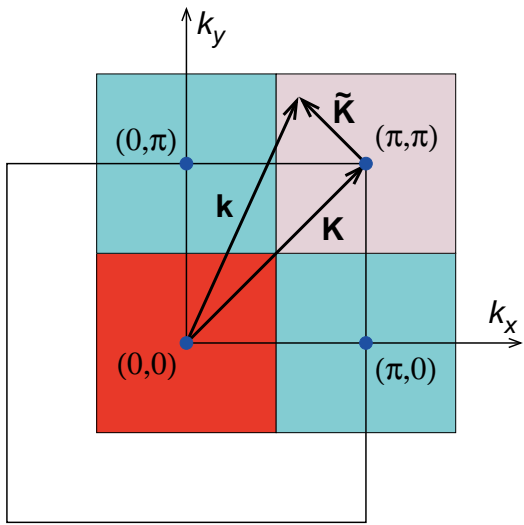}
\caption{\label{fig:bz} Partitioning of the square lattice Brillouin
  zone within the $N_c =4$ DCA method. There are four sectors
  characterized by the cluster momentum {\bf K}. An arbitrary
  reciprocal space vector {\bf k} is represented as a sum of {\bf K}
  and a vector $\tilde{\bf K}$ running within the cell labeled by {\bf
    K}.  }
\end{figure}

DCA is the cluster extension of single-site dynamical mean field
theory (DMFT)~\cite{Georges1996}, which includes, to a certain degree,
momentum dependencies.  Since the hopping matrix for sites within the
considered cluster and for sites on different clusters is the same, in
contrast to the cellular DMFT approach~\cite{Kotliar2001}, the DCA
self-consistent equation can be written in momentum space with the
assumption that the self-energy is constant in the Brillouin zone
sectors that are considered.  The DCA self-consistency equation is
given as
\begin{equation}\label{DCAE}
\overline{G}_{\sigma}({\bf K},i\omega_n)=\frac{1}{N}\sum_{\tilde{\bf K}}
\frac{1}{i\omega_n+\mu-\epsilon_{{\bf K+\tilde{K}}}-\Sigma_{\sigma}({\bf K},i\omega_n)},
\end{equation}
where $N$ is the number of $k$ points in each Brillouin zone sector (compare Figure~\ref{fig:bz}),
$\mu$ the chemical potential, ${\bf K}$ is the cluster momentum,
$\epsilon_{{\bf K+\tilde{K}}}$ the dispersion relation, $\omega_n$ are
the fermionic Matsubara frequencies, and where the summation over
${\bf \tilde{K}}$ is performed in each Brillouin zone sector. In our
calculations, we employed the DCA cluster with $N_c =4$, where ${\bf
  K} = (0,0),(0,\pi),(\pi,0)$, and $(\pi,\pi)$.  The converged
self-energy $\Sigma_{\sigma}({\bf K},i\omega_n)$ is evaluated by means
of Eq.~(\ref{DCAE}) and the Dyson equation.  We employ the interaction
expansion continuous-time quantum Monte Carlo approach as an impurity
solver~\cite{Rubtsov2005,Gull2011}. All calculations presented here
are for a temperature $T/t=0.05$ and we perform more than $10^{7}$ QMC
samplings for the impurity Green's function $G(i\omega_n)$, in order
to keep the QMC statistical errors smaller than $5 \times 10^{-3}$ for
the first Matsubara frequency.  As the Matsubara frequencies increase,
the error bar decreases.

\begin{figure}
\includegraphics[width=1.0\columnwidth]{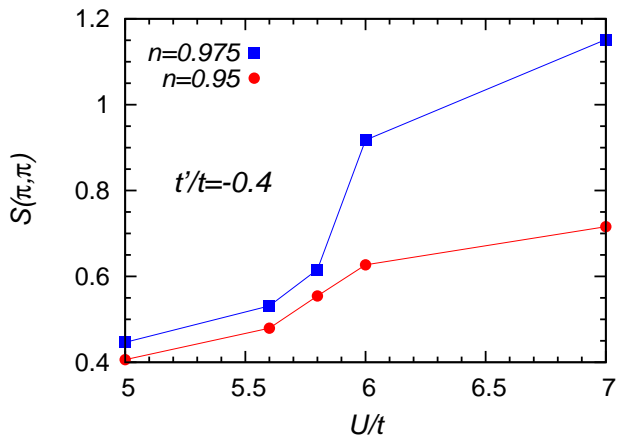}
\caption{\label{fig:Peak_Sq} (Color online) Evolution of the static
  structure factor correlations $S(q)=\langle s_{-q}s_q\rangle$ at
  $Q=(\pi,\pi)$ as a function of $U/t$ for $n=0.975$ (blue squares)
  and $n=0.95$ (red circles) at $t'/t=-0.4$. The appearance of the non-FL 
  region is characterized by a rapid increase in the
  short-range antiferromagnetic correlations. Data are obtained by
  means of VMC simulations on an $L=162$ lattice size.}
\end{figure}

\section{Energy and double occupancies}~\label{sec:energy}

In Fig.~\ref{fig:energy_com}, we compare the average number of double
occupancies per site $D=\langle n_{i,\uparrow}n_{i,\downarrow}\rangle$
and the energy, for $U/t=7.0$ and $t'=-0.4t$ as a function of the electron
density $n$, as obtained from VMC and DCA simulations. For the DCA
calculations, the energy is calculated by
\begin{equation}
E = \frac{T}{N} \sum_{n,\boldsymbol{k},\sigma} [\epsilon_{\boldsymbol{k}} 
\overline{G}_{\sigma} (\boldsymbol{k},i\omega_n)]e^{i\omega_n 0^{+}} + UD,
\label{energyDCA}
\end{equation} 
where the index ${\boldsymbol k}$ runs over the first Brillouin zone
and we considered the asymptotic behavior of the self-energy in the
limit of large Matsubara frequencies $i\omega_n$: $\Sigma_{\sigma}
({\bf K},i\omega_n) \sim U^2 n_{\sigma} (1-n_{\sigma}) / i\omega_n$.
The error $\Delta_n$ of the self-energy $\sum(i\omega_n)$ is calculated by
\begin{equation}
\Delta_n = \frac{2\xi}{G(i\omega_n)^2},
\end{equation}\label{eq:error}   
where $i\omega_n$ is the Matsubara frequency and $\xi$ is the error of $G(i\omega_n)$; 
it turns out that $\Delta_n$ is almost constant as a function of $n$. 
The error on the energy is then estimated as the difference between the energy 
calculated in Eq.~(\ref{energyDCA}) by using $\sum(i\omega_n)+\Delta_n$ as the self-energy 
and the energy calculated by using $\sum(i\omega_n)-\Delta_n$ as the self-energy.

For the large value of $U/t$ presented in Fig.~\ref{fig:energy_com}, both approaches show good
agreement in the double occupancy and in the energy, even if
the energies in DCA are characterized by large error bars, due to the high-frequency tail of the self-energy.

In Fig.~\ref{fig:Energy}, we present the double occupancy $D$ and the
energy close to half filling, for $n=0.95$ and $t'=-0.4t$, as a
function of the interaction strength $U/t$. We observe that
the results for the double occupancies at strong coupling ($U/t \sim 7.0$)
and at weak coupling ($U/t < 3.5$) are in good agreement
 but they differ in the range $U/t=4-6$. This discrepancy
is related to  the fact that VMC and DCA predict different values 
for the location of $U_c/t$ of the metal-insulator transition at half-filling.  
The singlet RVB state is dominant in the DCA on a $2 \times 2$ cluster, thus favoring 
correlated states, and also the VMC results can be slightly dependent on the accuracy 
of the trial wave function. 
The metal-insulator transition at half-filling, which is weakly first-order, takes place at
$U_c/t=5.8\pm0.2$ within VMC~\cite{tocchio_FS}, while for DCA, on a $2 \times 2$
cluster, the critical $U_c/t$ is approximately estimated at $U_c/t=4.6\pm0.4$~\cite{gullEuro,gull2}.
Note that the metal-insulator transition in VMC has been determined 
by looking at the static structure factor $N(q)=\langle n_{-q}n_q\rangle$, where 
$n_q=1/\sqrt{L}\sum_{r,\sigma} e^{iqr}n_{r,\sigma}$ is the Fourier transform 
of the particle density. Indeed, the metallic phase is characterized by $N(q)\sim q$ for $q\to 0$, 
which implies a vanishing gap for particle-hole excitations. On the contrary, 
in the insulating phase, $N(q)\sim q^2$ for $q \to 0$, 
implying the fact that the charge gap is finite.

At difference with the double occupancies, 
the energy data do not show in general good agreement within the two methods. 
One possible reason for the discrepancy is that 
the energy in DCA is calculated in terms of the self-energy, see Eq.~\ref{energyDCA},  
that is taken as constant within the Brillouin zone sectors. This means that we cannot obtain a true $k$-dependent 
self-energy in our small cluster DCA calculations. Furthermore, the high-frequency part 
of the self-energy is evaluated according to an approximate formula that may introduce 
a further systematic error in the calculated self-energy. Also VMC energies are not exact 
and further improvements in the wave function may lower the variational energy.

Signatures of the metal-insulator transition occurring at half filling
are visible at low doping in VMC calculations, where both the double occupancy $D$ and the energy show kinks for
interactions $U$ close to the critical $U_c$. Similar signatures are more difficult to 
infer from the DCA data around $U/t=4.5$ due to the larger error bars,
 but can be clearly observed in the behavior 
of the one-particle self-energy, as we discuss in the next section. In fact, both approaches, irrespective
of the actual location of the MIT predicted in each method, exhibit a rapid crossover 
between a weak-coupling Fermi-liquid
(FL) and an intermediate to large $U$ non-FL regime for small
but finite dopings. This conclusion will be discussed in the next section, based on the observed
behavior of the one-particle self-energy and of the two-particle
correlation functions.

\begin{figure}
\includegraphics[width=1.0\columnwidth]{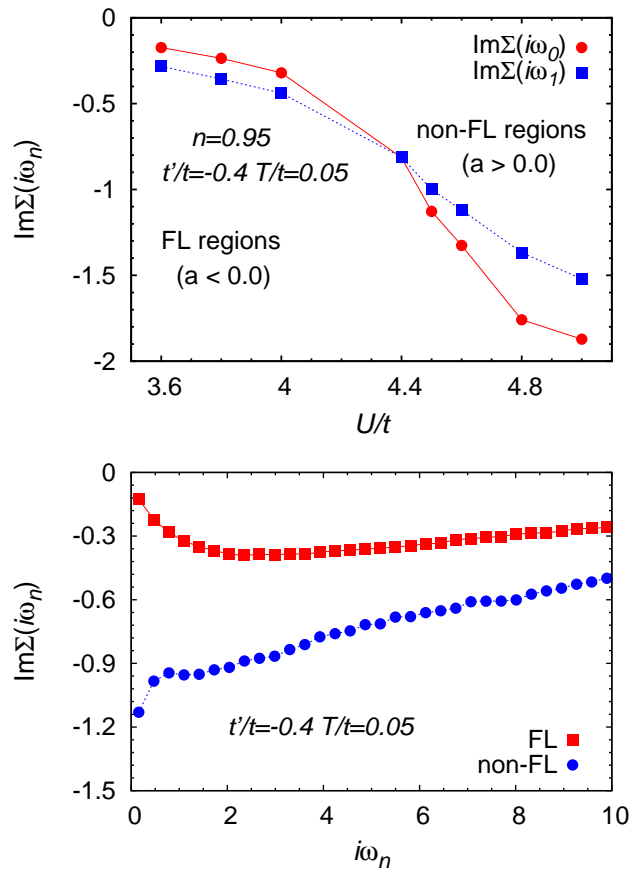}
\caption{\label{fig:DCA_selfenergy}
(Color online) Upper panel: The imaginary part of the lowest (red circles) and the second lowest 
(blue squares) self-energy values in Matsubara frequencies as a function of $U/t$ 
at $T/t=0.05$ with $n=0.95$ and $t'=-0.4t$. Data have been 
obtained  by DCA using a $2 \times 2$ cluster. The crossing 
point of both self-energies occurs around $U/t=4.5$, which is  
the same critical $U_c /t$ where the kink in the 
double occupancy data of Fig.~\ref{fig:Energy} is located. 
The quantity $a$ is defined in Eq.~(\ref{eq:slope}). 
Lower panel: The imaginary part of the self energy 
as a function of the Matsubara frequencies at $U/t=3.6$ and 
$n=0.94$ (FL, red squares) and at $U/t=5.0$ and $n=0.96$ 
(non-FL, blue circles). Data have been 
obtained  by DCA using a $2 \times 2$ cluster.}
\end{figure}

\section{Non-Fermi liquid vs. Fermi liquid properties}~\label{sec:NFL}

In this Section, we explore in more detail the FL and non-FL 
properties of the frustrated Hubbard model close to half filling,
using both the VMC and DCA approaches. Long-range static 
correlations can be evaluated within VMC but not within DCA,
due to cluster-size restrictions. Dynamical quantities like
the one-particle self-energy can however be calculated within
DCA and not with VMC; the two approaches complement each other
nicely.

\begin{figure}
\includegraphics[width=1.0\columnwidth]{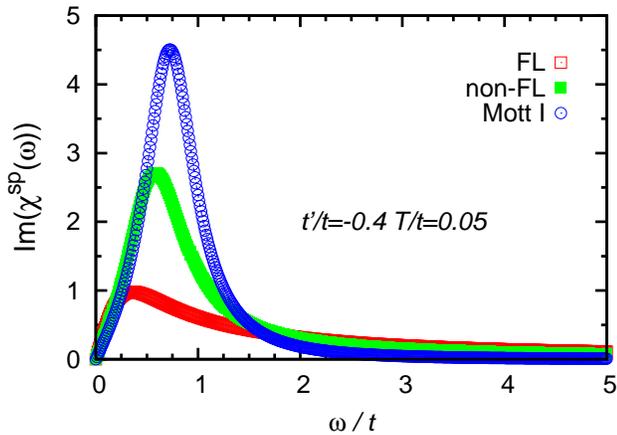}
\caption{\label{fig:DCA_twoparticle}
(Color online) The local dynamical spin susceptibility 
$\rm{Im} (\chi^{sp} (\omega))$ for 
$U/t=3.6$ and $n=0.94$ (FL, open red squares), 
$U/t=5.0$ and $n=0.96$ (non-FL, filled green squares) 
and $U/t=7.0$ and $n=1.0$ (Mott insulator, open blue circles),  
as a function of the real frequency $\omega /t$ at $T/t=0.05$ 
and $t'=-0.4t$. A Pad\'e approximation has been employed for 
the analytical continuation of $\chi^{sp} (\tau)$.}
\end{figure}

The nature of the non-Fermi liquid region has been characterized,
using VMC simulations~\cite{tocchio_FS},
by a strong renormalization of the underlying Fermi surface and a 
small violation of the Luttinger sum rule. Here, we assess the 
magnetic properties at low dopings via the static structure factor,
defined as 
\begin{equation}
S(q) = \frac{1}{L} \sum_{m,n} e^{i q (R_m-R_n)} 
\langle S_m^z S_n^z \rangle,
\end{equation}
where $S_m^z$ is the $z$-component of the spin operator on 
site $m$ and where $L$ denotes the total number of sites. 
The presence of (short-range) antiferromagnetic correlations 
is signaled by the appearance of a (non-diverging) peak in 
$S(q)$, located at $Q=(\pi,\pi)$. 

As shown in Fig.~\ref{fig:Peak_Sq}, for the two electron
densities $n=0.975$ and $n=0.95$ at $t'/t=-0.4$, the non-Fermi liquid state at 
$U/t \gtrsim 6$ is characterized by antiferromagnetic correlations 
which are substantially enhanced with respect to the weakly correlated 
metallic phase at $U/t \lesssim 6$. For the smaller doping,
$n=0.975$, the two regimes are clearly separated by a rapid increase
in the value of $S(Q)$, which could be compatible with a first-order 
transition, while for the larger doping the jump is less evident 
and the observed rise in the strength of the short-ranged 
spin-spin correlations is more in agreement with a smooth crossover.
 
Next, we plot the imaginary part of the lowest and the second lowest Matsubara frequency self-energy values  
as a function of $U/t$ at $T/t=0.05$, $n=0.95$ and $t'=-0.4t$ 
(see Fig.~\ref{fig:DCA_selfenergy}, upper panel), obtained using
DCA. We define 
\begin{equation}\label{eq:slope}
 a=\frac{{\rm Im}(\Sigma(i\omega_1)) - {\rm Im}(\Sigma(i\omega_0))}{\omega_1 - \omega_0}\,,
\end{equation}
where ${\rm Im}(\Sigma (i{\omega_0}))$ is the imaginary part of the lowest Matsubara frequency self-energy value
and ${\rm Im}(\Sigma (i{\omega_1}))$ is the imaginary part of the second lowest Matsubara self-energy value.
Negative ratios $a<0$ indicate (quasi)-FL behavior, 
while positive ratios $a>0$ suggest
a non-Fermi liquid state. Indeed, the value of the imaginary part of the self-energy 
converges to zero (or to small values due to the effect of temperature) for $i\omega_n\to 0$ in the Fermi-liquid regime, 
while it is monotonically decreasing for $i\omega_n\to 0$
in the non-Fermi liquid state, see Fig.~\ref{fig:DCA_selfenergy}, lower panel. 
From the data presented in
Fig.~\ref{fig:DCA_selfenergy}, we find a critical $U_c/t=4.5$ 
between FL and non-FL states, which is in agreement with the 
location of the small kink in the 
double occupancy $D$, compare Fig.~\ref{fig:Energy}.

\begin{figure}
\includegraphics[width=1.0\columnwidth]{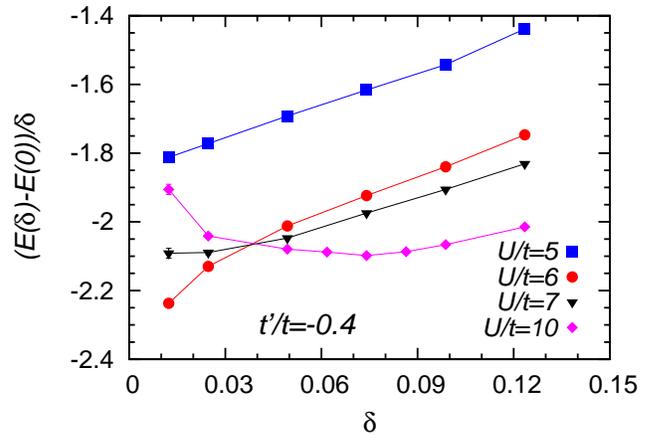}
\caption{\label{fig:Phase_separation}
(Color online) The hole energy 
$E_{\delta}(\delta)=(E(\delta)-E(0))/\delta$ 
as a function of doping for different values of $U/t$ 
at $t'/t=-0.4$. Data are obtained by means of VMC simulations on 
an $L=162$ lattice size}
\end{figure}
  
Finally, we present in Fig.~\ref{fig:DCA_twoparticle}
the local dynamical spin susceptibility 
$\rm{Im}(\chi^{sp} (\omega))$ obtained within DCA with $10^8$ QMC samplings, by performing
the  analytical continuation
of
\begin{equation}
 \chi^{sp} (\tau) = \langle S^z (\tau) S^z(0)\rangle\,,
\end{equation}
where $S^z = \frac{1}{2} (n_{\uparrow} - n_{\downarrow})$ and $\tau$
is the imaginary time, with the help of the Pad\'e approximation.  The
$\rm{Im} (\chi^{sp} (\omega))$ around $\omega /t =0.5$ is strongly
suppressed in the FL region at $U/t=3.6$ and $n=0.94$. This
low-frequency peak is dominant in the non-FL region at $U/t=5.0$ and
$n=0.96$ and in the Mott insulating region at $U/t=7.0$ and $n=1$. We
relate the enhancement of the low-energy peak in non-FL and Mott
insulator regions to the formation of short-range antiferromagnetic
correlations, as presented in Fig.~\ref{fig:Peak_Sq}, in terms of
increased RVB-type singlet pairing. These results demonstrate the
complementarity of the DCA and VMC methods
in identifying the region of possible non-FL behavior
 and they are compatible 
with studies of the Kagome and the triangular
lattice~\cite{Kyung2007,Ohashi2006}.

\section{Phase separation}~\label{sec:phase_sep}

We investigate now the possible occurrence of phase separation in the
hole-doped regime $n<1$, by considering the hole energy defined in
Eq.~(\ref{Emery}). The system is unstable to phase separation for
$\delta < \delta_c$, if the hole energy has a minimum at a critical
doping $\delta_c$, while a monotonically increasing hole energy
corresponds, on the other hand, to an energetically favorable
homogeneous solution for the doping levels investigated. Physically,
phase separation is driven by magnetic correlations, which can be
substantially increased, in the phase without holes, at the expense of
the kinetic energy.

\begin{figure}
\includegraphics[width=1.0\columnwidth]{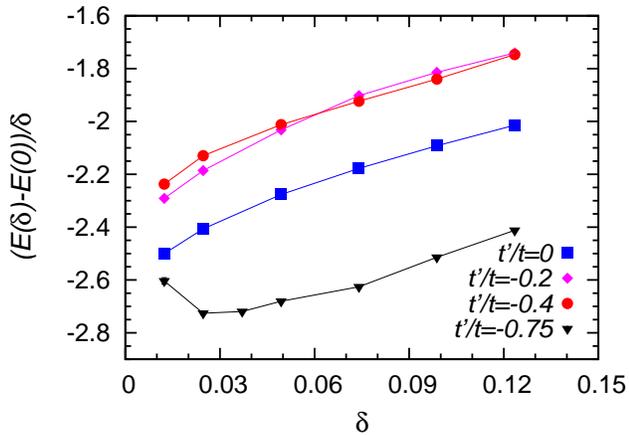}
\caption{\label{fig:Phase_sep_MIT}
(Color online) The hole energy 
$E_{\delta}(\delta)=(E(\delta)-E(0))/\delta$ 
as a function of doping for the four cases 
$U/t=6, t'/t=0$ (blue squares), 
$U/t=6, t'/t=-0.2$ (pink diamonds), 
$U/t=6, t'/t=-0.4$ (red circles) and 
$U/t=8, t'/t=-0.75$ (black triangles). 
The data has been obtained by means of VMC simulations
on an $L=162$ lattice.}
\end{figure}

In Fig.~\ref{fig:Phase_separation}, we focus on the case $t'/t=-0.4$,
and vary the interaction strength $U/t$. As expected, we find that 
no phase separation is possible when the half-filled case is 
metallic ($U/t=5$), due to the absence of a relevant magnetic 
energy scale. Phase separation is also not observed,
interestingly, when $U$ is only slightly above the critical 
interaction $U_c$ ($U/t=6$, in Fig.~\ref{fig:Phase_separation}), 
at least for small but finite doping levels $\delta \gtrsim 0.02$. 
At very low dopings $\delta \lesssim 0.02$ (and $U/t=7$) phase 
separation could possibly be present. However, since the model is expected to exhibit 
a magnetic instability close to half-filling, a possible phase separation occurring 
for $\delta \lesssim 0.02$ would be masked by long-range magnetic order. 
For larger values of $U/t$ 
phase separation seems however to be energetically favorable 
for a wider range $\delta\lesssim 0.08$ of doping levels. The
data for larger electron repulsion ($U/t\sim20$) are not shown 
in Fig.~\ref{fig:Phase_separation}, but they exhibit trends 
similar to the case $U/t=10$. However, we cannot exclude that better estimates 
of the ground-state energy can reduce this tendency to phase separate, as detailed 
in Ref.~\onlinecite{becca}. This is a consequence of the fact
that good estimates for the energy of low-doping states are hard to
obtain because of the strong local correlations.

In Fig.~\ref{fig:Phase_sep_MIT}, we show the hole energy as a function
of doping for four values of the next-nearest neighbor hopping,
ranging from $t'/t=0$ to $t'/t=-0.75$.  The value of the Coulomb
repulsion is slightly above the critical $U_c$, that is located
between $U/t=5$ and $U/t=6$ for $-0.4<t'/t<0$, while it is located
between $U/t=7$ and $U/t=8$ at $t'/t=-0.75$. In all the cases relevant
for the cuprates, $t'/t\simeq[-0.1,-0.4]$, no phase separation occurs.
Only in a small range $\delta \lesssim 0.02$ a tendency for phase
separation is observed for $t'/t=-0.75$. Thus, our data suggest that
for the cuprates
 the Hubbard model is not unstable 
against phase separation when the value of the electronic 
repulsion is chosen to be close to the critical interaction 
$U_c$ of the Mott-Hubbard transition.  

\section{Conclusions}~\label{sec:concl}

When investigating correlated electron systems numerical or analytical
approximations are generically necessary and the accuracy of the
employed approach is notoriously difficult to control. Here we compare
results obtained by two complementary approaches, DCA and VMC. We find
good agreement for the calculation of the double occupancies, apart from the value of the critical
Hubbard-$U$ for the Mott-Hubbard metal-insulator transition at half
filling, while comparing energies is more problematic, as discussed in Sec.~\ref{sec:energy}. 
We use the complementary information, static long-range
correlations provided by VMC and dynamical properties provided by DCA,
to investigate the physics at finite but low doping levels. We find
that the crossover from a weakly correlated electron state at
intermediate to small values of $U$ to a non-Fermi liquid state at
intermediate to large values of $U$ is characterized by a strong
increase in local magnetic correlations.  In this respect, we do not
find evidence for non-Fermi liquid properties below the critical
Hubbard-$U$, as suggested instead by a recent DCA
study~\cite{werner,gull2}.  Our result is a solid feature of the VMC
approach~\cite{tocchio_FS}. The
application of DCA in the present work has been oriented to supplement the
dynamical properties that are missing in VMC and we do not exclude
that other studies within DCA can show evidence of non-Fermi liquid properties
also below the critical $U$, though this is in contradiction
with VMC. This remains a controversial issue~\cite{gull2,dayal} and needs  further
study.
 In Sec.~\ref{sec:phase_sep}, we
investigate the stability of the non-Fermi liquid state      
against phase separation and find it to be stable for all parameters
relevant for the cuprates. These studies further prove
the synergies obtainable when using complementary methods for the study 
of frustrated and correlated electron systems.

\acknowledgments

We would like to thank the Deutsche Forschungsgemeinschaft for
financial support through grants SFB/TR49 (L.F.T.) and FOR 1346
(H.L.), and the Helmholtz association for financial support through
grant HA216/EMMI (H.O.J.). L.F.T. thanks Federico Becca for useful discussions.  



\begin{thebibliography}{99}

\bibitem{capone} M. Civelli, M. Capone, S. S. Kancharla, O. Parcollet, and 
G. Kotliar, {\prl} {\bf 95}, 106402 (2005).

\bibitem{Sakai2009} S. Sakai, Y. Motome, and M. Imada, {\prl} {\bf 102},
056404 (2009).
 
\bibitem{civelli} M. Civelli, {\prb} {\bf 79}, 195113 (2009).

\bibitem{jarrell} E. Khatami, K. Mikelsons, D. Galanakis, A. Macridin, J. Moreno, R. T. Scalettar, and M. Jarrell, 
{\prb} {\bf 81}, 201101(R) (2010).

\bibitem{gull} E. Gull, M. Ferrero, O. Parcollet, A. Georges, and A. J. Millis, 
{\prb} {\bf 82}, 155101 (2010).

\bibitem{sordi} G. Sordi, K. Haule, and A.-M. S. Tremblay, 
{\prb} {\bf 84}, 075161 (2011). 

\bibitem{liebsch} A. Liebsch and N.-H. Tong, Phys. Rev. B {\bf 80},
165126 (2009).

\bibitem{rice} K.-Y. Yang, T. M. Rice, and F.-C. Zhang, {\prb} {\bf 73}, 174501 
(2006).

\bibitem{markiewicz} R. S. Markiewicz, J. Lorenzana, G. Seibold, and A. Bansil,  
{\prb} {\bf 81}, 014509 (2010).

\bibitem{randeria} R. Sensarma, M. Randeria, and N. Trivedi, 
{\prl} {\bf 98}, 027004 (2007).

\bibitem{tocchio_FS} L. F. Tocchio, F. Becca, and C. Gros, 
{\prb} {\bf 86}, 035102 (2012).

\bibitem{werner} M. Sentef, P. Werner, E. Gull, and A. P. Kampf, 
{\prl} {\bf 107}, 126401 (2011). 

\bibitem{gull2} E. Gull, O. Parcollet, and A. J. Millis, 
arXiv:1207.2490 (unpublished).

\bibitem{sordi2} G. Sordi, P. S\'emon, K. Haule, and A.-M. S. Tremblay, 
{\prl} {\bf 108}, 216401 (2012).

\bibitem{anderson} P. W. Anderson, 
Science {\bf 235}, 1196 (1987).

\bibitem{emery} V. J. Emery, S. A. Kivelson, and H. Q. Lin, 
{\prl} {\bf 64}, 475 (1990).

\bibitem{becca} F. Becca, M. Capone, and S. Sorella, 
{\prb} {\bf 62}, 12700 (2000).

\bibitem{Maier} A. Macridin, M. Jarrell, and Th. Maier, 
{\prb} {\bf 74}, 085104 (2006).

\bibitem{aichhorn} M. Aichhorn, E. Arrigoni, M. Potthoff, and W. Hanke, 
{\prb} {\bf 76}, 224509 (2007).

\bibitem{chang} C.-C. Chang and S. Zhang, 
{\prb} {\bf 78}, 165101 (2008).

\bibitem{chen} K.-S. Chen, S. Pathak, S.-X. Yang, S.-Q. Su, D. Galanakis, K. Mikelsons, M. Jarrell, and J. Moreno, 
{\prb} {\bf 84}, 245107 (2011). 

\bibitem{trivedi} S. Y. Chang, S. Pathak, and N. Trivedi, 
{\pra} {\bf 85}, 013625 (2012).

\bibitem{chen2} K.-S. Chen, Z. Y. Meng, T. Pruschke, J. Moreno, and M. Jarrell, 
Phys.  Rev. B {\bf 86}, 165136 (2012).

\bibitem{watanabe} S. Watanabe and M. Imada, 
J. Phys. Soc. Jpn. {\bf 73}, 1251 (2004).

\bibitem{valenti} R. Valent\'i and C. Gros,
{\prl} {\bf 68}, 2402 (1992).

\bibitem{lugas} M. Lugas, L. Spanu, F. Becca, and S. Sorella,  
{\prb} {\bf 74}, 165122 (2006).

\bibitem{troyer} P. Corboz, S. R. White, G. Vidal, and M. Troyer, 
{\prb} {\bf 84}, 041108 (2011). 

\bibitem{hu} W.-J. Hu, F. Becca, and S. Sorella, 
{\prb} {\bf 85}, 081110(R) (2012). 

\bibitem{backflow} L. F. Tocchio, F. Becca, and C. Gros, 
{\prb} {\bf 83}, 195138 (2011).


\bibitem{Hettler1998} M. H. Hettler, A. N. Tahvildar-Zadeh, M. Jarrell, T. Pruschke, and H. R. Krishnamurthy, 
{\prb} {\bf 58}, 7475(R) (1998).


\bibitem{Maier2005} T. Maier, M. Jarrell, T. Pruschke, and M. Hettler,
Rev. Mod. Phys {\bf 77}, 1027 (2005).

\bibitem{lanczos}
   L. F. Tocchio, F. Becca, A. Parola, and S. Sorella,
{\prb} {\bf 78}, 041101(R) (2008).

\bibitem{andersen} E. Pavarini, I. Dasgupta, T. Saha-Dasgupta, O. Jepsen, and O. K. Andersen, 
{\prl} {\bf 87}, 047003 (2001).

\bibitem{capello} M. Capello, F. Becca, S. Yunoki, 
                  M. Fabrizio, and S. Sorella,
{\prb} {\bf 72}, 085121 (2005).

\bibitem{grosbcs} C. Gros, {\prb} {\bf 38}, 931(R) (1988).

\bibitem{zhang} F. C. Zhang, C. Gros, T. M. Rice, and H. Shiba, 
Supercond. Sci. Technol. {\bf 1}, 36 (1988).

\bibitem{grosAnnals} C. Gros,
Annals of Physics {\bf 189}, 53 (1989).

\bibitem{stoc_ref} S. Yunoki and S. Sorella, 
Phys. Rev. B {\bf 74}, 014408 (2006).

\bibitem{Georges1996} A. Georges, G. Kotliar, W. Krauth, and M. J. Rozenberg,
Rev. Mod. Phys {\bf 68}, 13 (1996).

\bibitem{Kotliar2001} G. Kotliar, S. Y. Savrasov, G. Palsson, and G. Biroli, 
{\prl} {\bf 87}, 186401 (2001).

\bibitem{Rubtsov2005} A. N. Rubtsov, V. V. Savkin, and A. I. Lichtenstein,
{\prb} {\bf 72}, 035122 (2005).

\bibitem{Gull2011} E. Gull, A. J. Millis, A. I. Lichtenstein, A. N. Rubtsov, M. Troyer and P. Werner,
Rev. Mod. Phys. {\bf 83}, 349 (2011).


\bibitem{inversion} E. Gull, P. Werner, A. Millis, and M. Troyer, 
{\prb} {\bf 76}, 235123 (2007).

\bibitem{gullEuro} E. Gull, P. Werner, X. Wang, M. Troyer, and A. J. Millis,
  Europhys. Lett. {\bf 84}, 37009 (2008).

\bibitem{Kyung2007} B. Kyung,
{\prb} {\bf 75}, 033102 (2007).

\bibitem{Ohashi2006} T. Ohashi, N. Kawakami, and H. Tsunetsugu, 
{\prl} {\bf 97}, 066401 (2006).

\bibitem{dayal} S. Dayal, R. T. Clay, and S. Mazumdar, {\prb} {\bf 85}, 165141
                (2012).

\end{thebibliography}
\end{document}